\documentstyle[epsfig]{article}
\font\tenbf=cmbx10
\font\tenrm=cmr10
\font\tenit=cmti10
\font\elevenbf=cmbx10 scaled\magstep 1
\font\elevenrm=cmr10 scaled\magstep 1

\renewenvironment{thebibliography}[1]
 { \tenrm
 \baselineskip=10pt
   \begin{list}{\arabic{enumi}.}
    {\usecounter{enumi} \setlength{\parsep}{0pt}
     \setlength{\itemsep}{3pt} \settowidth{\labelwidth}{#1.}
     \sloppy
    }}{\end{list}}

\begin{document}
\begin{center}{\tenbf
EXOTIC BARYONS AND MULTIBARYONS IN CHIRAL SOLITON MODELS. 
\\}
\vglue 0.5cm
{\tenrm V. KOPELIOVICH \\}
{\tenit Institute for Nuclear Research of RAS, Moscow 117312, Russia\\
}
\vglue 0.2cm
\end{center}
{\rightskip=3pc
\leftskip=3pc
\tenrm\baselineskip=10pt
\noindent
Recently observed baryonic resonance with positive strangeness is discussed.
Mass and width of this 
resonance are in agreement with the chiral soliton model predictions. A number 
of other exotic states are predicted within this approach, some of them are
probably observed in experiments. Existence of exotic multibaryons is expected
as well, with positive strangeness or beauty, and negative charm. The 
possibility of binding of heavy anti-flavor is noted.
 \vglue 0.2cm}
\baselineskip=13pt
\elevenrm
\section{Introduction}
In recent experiments \cite{1,2,3} the baryonic resonance has been discovered
with positive strangeness and rather small width, $\Gamma\, <\, 24\;Mev$, 
and subsequent experiment \cite{4} has confirmed
this discovery \footnote{These data, as well as \cite{5} where resonance was 
observed in analysis of neutrino/antineutrino interactions with nuclei, 
became available after the symposium.}.
This resonance is observed independently in different reactions on different
experimental setups in Japan, Russia, USA and FRG, therefore only few doubts
remain now that it really exists.

This baryon, predicted theoretically in \cite{6,7,8} originally called 
$Z^+$\cite{8} and later $\Theta^+$, together with the well known
resonances $\Lambda(1520)$ and $\Xi (1530)$ has one of the smallest widths
among available baryon resonances. It has necessarily one quark-antiquark
pair in its wave function since baryons made of 3 valence quarks only can have
negative strangeness, $S\,<0$.

Besides this, some hints have been obtained on detector CLAS in reaction of
$\pi^+\pi^-$ electroproduction on protons for existence of new resonance with 
zero strangeness, positive parity, strong coupling to the $\Delta\pi$ channel
and weak to the $N\rho$ \cite{9}. This resonance could belong to one of the
multiplets of exotic baryons considered in \cite{10}. Review of experimental 
situation, methods of detection of exotic and so called cryptoexotic states
(states with hidden exotics) before discovery of $\Theta^+$ can be found,
for example, in \cite{11}.
\section{Multiplets of exotic baryons} 
Exotic, in specific meaning of this word, are baryonic states which cannot 
be made of $3B$ valence quarks ($B$ is the baryon number) and should contain 
one or more quark-antiquark pairs.
Obviously, any state with positive strangeness is exotic one, as well as states
with large enough negative strangeness, $S<\; -3B$. Besides, for any value of 
hypercharge $Y$ or strangeness $S<0$ there are exotic states with large enough 
isospin, $I>(3B+S)/2$. It is due to the fact that nonzero isospin have only
nonstrange quarks $u,d$, and the number of nonstrange valence quarks equals to
$3B+S$.  The new-found hyperon with positive strangeness and at least one 
quark-antiquark pair in the wave function is called also the pentaquark state.
It is well known that baryons (hadrons, more generally) contain the 
so-called sea quarks and gluons which carry large fraction of their momenta.
But in the $\Theta$-pentaquark the $q\bar{q}$-pair has definite quantum number,
antistrangeness, therefore it is in fact valence quark-antiquark pair.

From theoretical point of view the existence of such states was not unexpected.
Such possibility was pointed out by a number of people within the quark models
\cite{12}, as well as in the chiral soliton approach \cite{13,14}. Analysis of
peculiarities of exotic baryons spectra, for arbitrary $B$-numbers, and 
estimates of energies for exotic $SU(3)$ multiplets was made in \cite{15}. 
First numerical estimates of the masses of the antidecuplet components were 
made in \cite{6,14,7}.
Relatively small mass of the components of antidecuplet, in particular 
$\Theta^+$, was predicted in a number of papers \cite{6,14,7,8,16}, and strictly 
speaking, it was not enough grounds for this in \cite{6}a,\cite{14}, 
because the mass 
splitting in the octet and decuplet of baryons was not described in these
papers. In the paper \cite{8} an assumption was important to provide the
prediction $M_{\Theta^+} = 1530 \;Mev$, that the nucleon resonance $N^*(1.71)$
is the nonstrange component of the antidecuplet. The small width $\Gamma_\Theta
\simeq 15\;Mev$ was obtained in \cite{8} only.

Topological soliton models are very economical and effective in predicting the
spectra of baryons and baryonic systems with various quantum numbers. The 
relativistic many-body problem to find the bound states in a system of three,
five, etc. quarks and antiquarks is not solved in this way, of course.
However, many unresolved questions of principle are circumvented so
that calculations of spectra of baryonic states become possible without
detalization of their internal structure.
In such models baryons or baryonic systems (nuclei) appear as quantized 
classical (chiral) fields configurations obtained in the procedure of classical
energy or mass minimization. Here important role plays the quantization 
condition \cite{17}
$$ Y_R \;=\;N_CB/3 \eqno(1) $$
where $Y_R$ is the "right" hypercharge, or hypercharge of the state in the 
body-fixed system, $N_C$ - the number of colors of underlying $QCD$, $B$ is 
baryon number coinciding with topological number characterizing the classical
field configuration. For each $SU(3)$ multiplet $(p,q)$ the maximal hypercharge
or triality $Y_{max}=(p+2q)/3$, and relation should be fulfilled evidently
$Y_{max} \geq Y_R$, or
$${p+2q\over 3} \geq {N_CB\over 3}, \eqno(2) $$
which means that
$$p+2q = 3(B+m) \eqno (3) $$
at $N_C=3$, with $m$ being positive integer. This quantization condition has
simple physical interpretation: we start from originally nonstrange configuration
which remains nonstrange in the body fixed system. All other components of the
$(p,q) \; SU(3)$ multiplet in the laboratory frame appear as a result of
rotation of this configuration in $SU(3)$ configuration space, are described by
Wigner final $SU(3)$-rotations functions, and each multiplet should contain 
original nonstrange state. It is natural to call the multiplets with $m=0$
the minimal multiplets \cite{15}, for $B=1$ the minimal multiplets are well 
known octet and decuplet, multiplets with smaller dimension are forbidden due
to Guadagnini quantization condition \cite{17} (recall that the number of 
components of the multiplet $N(p,q)=(p+1)(q+1)(p+q+2)/2$).

The states with $m=1$ contain at least one additional quark-antiquark pair.
Indeed, the maximal hypercharge $Y_{max}=2$ in this case, or strangeness
$S=+1$ for the upper components of such multiplets, i.e. the pair $q\bar{s}$
should be present in the wave function, $q=u$ or $d$. Due to $SU(3)$ invariance 
of strong interactions all other components of such multiplet should contain
additional quark-antiquark pair \cite{15}. One more restriction appears from 
the consideration of the isospin, really the components with maximal isospin.
It is easily to check, that $\{\overline{10}\}$, $\{27\}$ and $\{35\}$-multiplets
are the pentaquark states, but the multiplet with maximal $p$, $\{28\}$-plet
with $(p,q)=(6,0)$ contains already 2 $q\bar{q}$ pairs, i.e. it is septuquark.
This follows from the fact that this multiplet contains the state with $S=-5$
and the state with $S=1$, isospin $I=3$.
All baryonic multiplets with $B=m=1$ are shown in {\bf Fig.1}. 
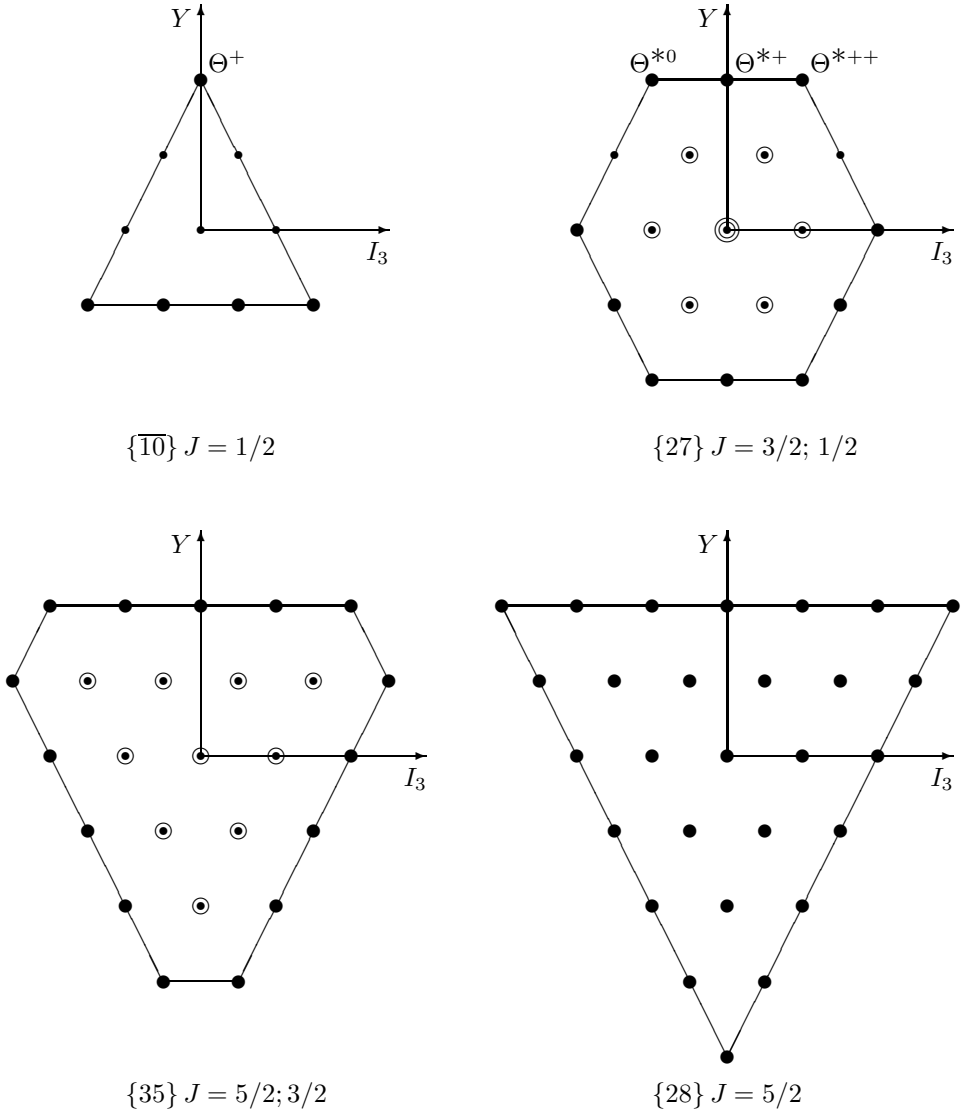
\begin{figure}[h]
\label{multiplet}
\setlength{\unitlength}{1.0cm}
\begin{flushleft}
\begin{picture}(12,14)
\put(3,11){\vector(1,0){2.5}}
\put(3,11){\vector(0,1){3}}
\put(2.6,13.7){$Y$}
\put(5.2,10.6){$I_3$}
\put(2,8){$\{\overline {10}\}\, J=1/2$}
\put(3.1,13.1){$\Theta^+$}

\put(3,13){\circle*{0.18}}
\put(2.5,12){\circle*{0.1}}
\put(3.5,12){\circle*{0.1}}
\put(2,11){\circle*{0.1}}
\put(3,11){\circle*{0.1}}
\put(4,11){\circle*{0.1}}
\put(1.5,10){\circle*{0.18}}
\put(2.5,10){\circle*{0.18}}
\put(3.5,10){\circle*{0.18}}
\put(4.5,10){\circle*{0.18}}

\put(1.5,10){\line(1,0){3}}
\put(1.5,10){\line(1,2){1.5}}
\put(4.5,10){\line(-1,2){1.5}}


\put(10,11){\vector(1,0){3}}
\put(10,11){\vector(0,1){3}}
\put(9.6,13.7){$Y$}
\put(12.7,10.6){$I_3$}
\put(9,8){$\{27\}\, J=3/2;\,1/2$}
\put(8.7,13.1){$\Theta$*$^0$}
\put(10.1,13.1){$\Theta$*$^+$}
\put(11.1,13.1){$\Theta$*$^{++}$}

\put(9,13){\circle*{0.18}}
\put(10,13){\circle*{0.18}}
\put(11,13){\circle*{0.18}}

\put(8.5,12){\circle*{0.1}}
\put(9.5,12){\circle*{0.1}}
\put(9.5,12){\circle {0.2}}
\put(10.5,12){\circle*{0.1}}
\put(10.5,12){\circle {0.2}}
\put(11.5,12){\circle*{0.1}}

\put(8,11){\circle*{0.18}}
\put(9,11){\circle*{0.1}}
\put(10,11){\circle*{0.1}}
\put(11,11){\circle*{0.1}}
\put(12,11){\circle*{0.18}}
\put(9,11){\circle {0.2}}
\put(10,11){\circle {0.2}}
\put(11,11){\circle {0.2}}
\put(10,11){\circle {0.3}}

\put(8.5,10){\circle*{0.18}}
\put(9.5,10){\circle*{0.1}}
\put(9.5,10){\circle {0.2}}
\put(10.5,10){\circle*{0.1}}
\put(10.5,10){\circle {0.2}}
\put(11.5,10){\circle*{0.18}}

\put(9,9){\circle*{0.18}}
\put(10,9){\circle*{0.18}}
\put(11,9){\circle*{0.18}}

\put(8,11){\line(1,2){1}}
\put(8,11){\line(1,-2){1}}
\put(9,13){\line(1,0){2}}
\put(9,9){\line(1,0){2}}
\put(12,11){\line(-1,2){1}}
\put(12,11){\line(-1,-2){1}}


\put(3,4){\vector(1,0){3}}
\put(3,4){\vector(0,1){3}}
\put(2.6,6.7){$Y$}
\put(5.7,3.6){$I_3$}
\put(2,-0.6){$\{35\}\, J=5/2;3/2$}

\put(1,6){\circle*{0.18}}
\put(2,6){\circle*{0.18}}
\put(3,6){\circle*{0.18}}
\put(4,6){\circle*{0.18}}
\put(5,6){\circle*{0.18}}

\put(0.5,5){\circle*{0.18}}
\put(1.5,5){\circle*{0.1}}
\put(1.5,5){\circle {0.2}}
\put(2.5,5){\circle*{0.1}}
\put(2.5,5){\circle {0.2}}
\put(3.5,5){\circle*{0.1}}
\put(3.5,5){\circle {0.2}}
\put(4.5,5){\circle*{0.1}}
\put(4.5,5){\circle {0.2}}
\put(5.5,5){\circle*{0.18}}

\put(1,4){\circle*{0.18}}
\put(2,4){\circle*{0.1}}
\put(3,4){\circle*{0.1}}
\put(4,4){\circle*{0.1}}
\put(5,4){\circle*{0.18}}
\put(2,4){\circle {0.2}}
\put(3,4){\circle {0.2}}
\put(4,4){\circle {0.2}}

\put(1.5,3){\circle*{0.18}}
\put(2.5,3){\circle*{0.1}}
\put(2.5,3){\circle {0.2}}
\put(3.5,3){\circle*{0.1}}
\put(3.5,3){\circle {0.2}}
\put(4.5,3){\circle*{0.18}}

\put(2,2){\circle*{0.18}}
\put(3,2){\circle*{0.1}}
\put(3,2){\circle {0.2}}
\put(4,2){\circle*{0.18}}

\put(2.5,1){\circle*{0.18}}
\put(3.5,1){\circle*{0.18}}

\put(0.5,5){\line(1,2){0.5}}
\put(0.5,5){\line(1,-2){2}}
\put(1,6){\line(1,0){4}}
\put(2.5,1){\line(1,0){1}}
\put(5.5,5){\line(-1,2){0.5}}
\put(5.5,5){\line(-1,-2){2}}


\put(10,4){\vector(1,0){3}}
\put(10,4){\vector(0,1){3}}
\put(9.6,6.7){$Y$}
\put(12.7,3.6){$I_3$}
\put(9,-0.6){$\{28\}\, J=5/2 $}

\put(7,6){\circle*{0.18}}
\put(8,6){\circle*{0.18}}
\put(9,6){\circle*{0.18}}
\put(10,6){\circle*{0.18}}
\put(11,6){\circle*{0.18}}
\put(12,6){\circle*{0.18}}
\put(13,6){\circle*{0.18}}

\put(7.5,5){\circle*{0.18}}
\put(8.5,5){\circle*{0.18}}
\put(9.5,5){\circle*{0.18}}
\put(10.5,5){\circle*{0.18}}
\put(11.5,5){\circle*{0.18}}
\put(12.5,5){\circle*{0.18}}

\put(8,4){\circle*{0.18}}
\put(9,4){\circle*{0.18}}
\put(10,4){\circle*{0.18}}
\put(11,4){\circle*{0.18}}
\put(12,4){\circle*{0.18}}

\put(8.5,3){\circle*{0.18}}
\put(9.5,3){\circle*{0.18}}
\put(10.5,3){\circle*{0.18}}
\put(11.5,3){\circle*{0.18}}

\put(9,2){\circle*{0.18}}
\put(10,2){\circle*{0.18}}
\put(11,2){\circle*{0.18}}

\put(9.5,1){\circle*{0.18}}
\put(10.5,1){\circle*{0.18}}
\put(10,0){\circle*{0.18}}
\put(7,6){\line(1,0){6}}
\put(10,0){\line(-1,2){3}}
\put(13,6){\line(-1,-2){3}}

\end{picture}
\vglue 0.7cm
\caption{The $I_3-Y$ diagrams for the baryon multiplets with $B=1,\;m=1$.
Large full circles show the exotic states, smaller - the cryptoexotic states
which can mix with nonexotic states from octet and decuplet.}

\end{flushleft}
\end{figure}

The minimal value of hypercharge is $Y_{min}=-(2p+q)/3$, the maximal isospin
$I_{max}=(p+q)/2$ at $Y=(p-q)/3$. Such multiplets as $\{27\}$, $\{35\}$
for $m=1$ and all multiplets for $m=2$, except the last one with $(p,q)=(9,0)$
in their internal points contain 2 or more states with different values of spin
$J$ (shown by double or triple circles in {\bf Fig.1}).
\section{The mass formula}
In the collective coordinates quantization procedure one introduces the angular
velocities of rotation of skyrmion in the $SU(3)$ configuration space, 
$\omega_k$, $k=1,...8$: $A^\dagger(t)\dot{A}(t) = -i \omega_k\lambda_k/2,\;
\lambda_k$ being Gell-Mann matrices, the collective coordinates matrix $A(t)$
is written usually in the form $A=A_{SU2}\,exp(i\nu\lambda_4)A'_{SU2}\,
exp(i\rho \lambda_8/\sqrt{3})$.
The corresponding contribution to the lagrangian is quadratic form in these 
angular velocities, with momenta of inertia, isotopical (pionic) $\Theta_\pi$ 
and flavor, or kaonic $\Theta_K$ as coefficients \cite{17}:
$$ L_{rot} = {1\over 2}\Theta_\pi (\omega_1^2+\omega_2^2+\omega_3^2) + 
{1\over 2}\Theta_K (\omega_4^2+...+\omega_7^2) - {N_cB \over 2\sqrt{3}} 
\omega_8. \eqno (4) $$
The expressions for these moments of inertia as functions of skyrmion profile
are presented below. The quantization condition $(1)$ discussed above follows
from the presence of linear in angular velocity $\omega_8$ term in $(4)$
originated from the Wess-Zumino-Witten term in the action of the model 
\cite{18}. 

The hamiltonian of the model can be obtained from $(4)$ by means of canonical 
quantization procedure \cite{17}:
$$ H = M_{cl} + {1\over 2\Theta_\pi} \vec{R}^2 + {1\over 2\Theta_K} 
\biggl[C_2(SU_3) -\vec{R}^2 -{N_c^2B^2\over 12} \biggr], \eqno (5)$$
where the second order Casimir operator for the $SU(3)$ group, 
$C_2(SU_3)=\sum_{a=1}^8 R_a^2$, with eigenvalues for the  $(p,q)$ multiplets
$C_2(SU_3)_{p,q}=(p^2+pq+q^2)/3 +p+q, $
for the $SU(2)$ group,
$C_2(SU2)=\vec{R}^2 =R_1^2+R^2_2+R^2_3= J(J+1) = I_R(I_R+1)$.

The operators $R_\alpha = \partial L/\partial\omega_\alpha$ satisfy definite
commutation relations which are generalization of the angular momentum 
commutation relations to the $SU(3)$ case \cite{17}. Evidently, the linear in
$\omega$ terms in lagrangian $(4)$ are cancelled in hamiltonian $(5)$.
The equality of angular momentum (spin) $J$ and the so called right or body 
fixed isospin $I_R$ used in $(5)$ takes place only for configurations of the
"hedgehog" type when usual space and isospace rotations are equivalent. This
equality is absent for configurations which provide the minimum of classical
energy for greater baryon numbers, $B\geq 2$.

For minimal multiplets $(m=0)$ the right isospin $I_R=p/2$, and it is easy to
check that coefficient of $1/2\Theta_K$ in $(5)$ equals to
$$\,K=\,C_2(SU_3)-\vec{R}^2-N_C^2B^2/12 \,=\,N_CB/2, \eqno (6) $$
for arbitrary $N_C$ \footnote{It should be kept in mind that for $N_C$ 
different from 3 the minimal multiplets for baryons differ from octet 
and decuplet. They have $(p,q)=(1,(N_C-1)/2),\; (3,(N_C-3)/2),...,\,(N_C,0)$. }.
So, $K$ is the same for all multiplets with $m=0$ \cite{15}, see {\bf Table 1}-
the property known long ago for the $B=1$ case \cite{17}.
For nonminimal multiplets there are additional contributions to the energy
proportional to $m/\Theta_K$ and $m^2/\Theta_K$, according to $(5)$\cite{15}. 
It means 
that in the framework of chiral soliton approach the "weight" of quark-
antiquark pair is defined by parameter $1/\Theta_K$, and this property of 
such models deserves better understanding.
\begin{center}
\begin{tabular}{|l|l|l|l|l|l|l|}
\hline
$(p,q)$& $N(p,q)$         &m &$C_2(SU_3)$&$J=I_R$ &$K(J_{max})$&$K(J_{max}-1)$\\
\hline
$(1,1)$&$\{8\}$            &0 & 3       &1/2  &3/2 &\\
$(3,0)$&$\{10\}$           &0 & 6       &3/2  &3/2 &\\
\hline
$(0,3)$&$\{\overline{10}\}$ &1&6         &1/2       &3/2+3&\\
$(2,2)$&$\{27\}$            &1&8         &3/2; 1/2  &3/2+2& 3/2+5\\
$(4,1)$&$\{35\}$            &1&12        &5/2; 3/2  &3/2+1& 3/2+6\\
$(6,0)$&$\{28\}$            &1&18        &5/2       &3/2+7&\\
\hline
$(1,4)$&$\{\overline{35}\}$ &2&12        &3/2; 1/2      &3/2+6& 3/2+9\\
$(3,3)$&$\{64\}$            &2&15        &5/2; 3/2; 1/2 &3/2+4& 3/2+9\\
$(5,2)$&$\{81\}$            &2&20        &7/2; 5/2; 3/2 &3/2+2& 3/2+9 \\
$(7,1)$&$\{80\}$            &2&27        &7/2; 5/2       &3/2+9& 3/2+16\\
$(9,0)$&$\{55\}$            &2&36        &7/2            &3/2+18& \\
\hline
\end{tabular}
\end{center}

{\bf Table 1.}{\tenrm The values of $N(p,q)$, Casimir operator $C_2(SU_3)$, spin
$J=I_R$, coefficient $K$ for first two values of $J$ for minimal $(m=0)$ and 
nonminimal $(m=1,\;2)$ multiplets of baryons.}
\vglue 0.2cm
It follows from {\bf Table 1} that for each nonzero $m$ the coefficient 
$K(J_{max})$ decreases with increasing $N(p,q)$, e.g. $K_{5/2}(35)\,<K_{3/2}(27)
\,<\,K_{1/2}(\overline{10})$. The following differences of the rotation energy
can be obtained easily:
$$ M_{10} - M_{8} = {3\over 2\Theta_\pi}. \eqno (7) $$
This relation is known since 1984 \cite{17}.
$$ M_{\bar{10}} - M_{8} = {3\over 2\Theta_K}, \eqno (8)$$
as it was stressed in \cite{8},
$$ M_{27,J=3/2}- M_{10} = {1\over \Theta_K}, \eqno(9) $$
$$ M_{27,J=3/2} - M_{\bar{10}} = {3\over 2\Theta_\pi} - {1\over 2\Theta_K},
 \eqno(10) $$
$$ M_{35,J=5/2} - M_{27,J=3/2} = {5\over 2\Theta_\pi} - {1\over 2\Theta_K}.
 \eqno(11) $$
If the relation took place $\Theta_K \ll \Theta_\pi$ then $\{27\}$-plet would 
be lighter than antidecuplet, and $\{35\}$-plet would be lighter than $\{27\}$.
In realistic case $\Theta_K$ is approximately twice smaller than $\Theta_\pi$
(see {\bf Table 2}, next section), and therefore the components of antidecuplet are lighter
than components of $\{27\}$ with same values of strangeness.
Beginning with some values of $N(p,q)$ coefficient $K$ increases strongly, as can
be seen from {\bf Table 1}, and this corresponds to the increase of the number 
of quark-antiquark pairs by another unity. The states with $J\,<\,J_{max}$ have
the energy considerably greater than that of $J_{max}$ states, by this reason 
they could contain also greater amount of $q\bar{q}$-pairs.

The formula $(5)$ is obtained in the rigid rotator approximation which is 
valid if the profile function of the skyrmion and therefore its dimensions and 
other properties are not changed when it is rotated in the configuration space.
It is necessary for this, that the rotation time in the configuration space,
$\tau_{rot}$ is smaller than the time of its deformation $\tau_{deform}$ 
under influence of the
forces due to presence of the terms in lagrangian violating the flavor symmetry,
i.e. $m_k/m_\pi >1,\; F_K/F_\pi >1$, see also next Section. Rotation time can be 
estimated easily, $\tau_{rot} \sim \pi/\omega$ with $\omega \sim \sqrt{C_2(SU3)}
/\Theta_K$. It is more difficult to estimate $\tau_{deform}$, one can state only
that it is greater than the time needed for light to cross the skyrmion,
$\tau_{travel} \sim 2R_H$. So, the rigid rotator approximation is valid if
$\pi \Theta_K \ll 2R_H\sqrt{C_2(SU3)}$. Numerically $\pi\Theta_K \simeq 
8\,Gev^{-1}$ and $2R_H\sqrt{C_2(SU3)}\simeq 12\,Gev^{-1}$ for decuplet and 
antidecuplet of baryons.

The alternative is the "soft" or slow rotator approximation when it is assumed 
that for each value of the angle of rotation in "strange" direction $\nu$ there
is enough time for the soliton to be deformed under influence of the flavor 
symmetry breaking forces \cite{19}. The realistic case is intermediate one,
but for the baryons the rigid rotator approach is more justified, due to above
estimate. With increasing $B$-number the slow rotator approach becomes more
actual. The dependence of the moments of inertia on $\nu$ is given by following
expressions \cite{19,20}:
$$ \Theta_K(\nu) = {1 \over 8} \int (1-c_f)\Biggl\{ F_K^2\Biggl(1\,-\;
\frac{2-c_f}{2}s_\nu^2\Biggr)+\,F_\pi^2\frac{2-c_f}{2}s_\nu^2 + {1 \over e^2}
 \Biggl[ f'^2 +{2s_f^2\over r^2} \Biggr] \Biggr\} 
d^3\vec{r}, \eqno (12) $$
$$\Theta_\pi(\nu) ={1\over 6}\int s_f^2\Biggl[F_\pi^2+(F_K^2-F_\pi^2)
c_fs_\nu^2 +{4\over e^2}\Biggl(f'^2+{s_f^2\over r^2}\biggr)\Biggr]
d^3\vec{r} \eqno(13) $$
These formulas hold for configurations of hedgehog type described by one 
profile function $f$, $c_f=cos\,f,\; s_f=sin\,f;\; s_\nu =sin\,\nu$.
$\Theta_K(\nu)$ decreases and $\Theta_\pi(\nu)$ increases with increasing $\nu$.
Rigid rotator approximation corresponds to $\nu=0$ since we start from 
nonstrange $SU(2)$-skyrmion. The decay constants $F_\pi,\;F_K$ are taken from
experiment: $F_\pi\simeq 186\,Mev$; the model parameter (Skyrme constant) $e$
is close to $4$. The dependence on $F_K$ in $(12,13)$ appears due
to nonadiabatic (time dependent) terms in the lagrangian which can have also 
other manifestations.
\section{Spectrum of baryonic states}
Expressions $(5),\;(6)$ and numbers given in {\bf Table 1} are sufficient 
to calculate the spectrum of baryons without mass splitting inside of $SU(3)$-
multiplets, as it was made e.g. in \cite{14,15}.
The mass splitting due to the presence of flavor symmetry breaking terms plays a 
very substantial role \cite{17,7,10}:
$$H_{SB}=\frac{1-D_{88}^{(8)}}{2}\Gamma_{SB} \eqno(14) $$
where the $SU(3)$ rotation function $D_{88}^8(\nu) =1-3s^2_\nu/2$,
$$\Gamma_{SB}={2\over 3}\Biggl[\Biggl({F_K^2\over F_\pi^2}m_K^2 -m_\pi^2\Biggr)
\Sigma +(F_K^2-F_\pi^2)\tilde{\Sigma}\Biggr] \eqno (15) $$
$$ \Sigma = \frac{F_{\pi}^2}{2} \int (1-c_f) d^3\vec{r}, $$
$$\tilde{\Sigma}= {1\over 4}\int c_f \Biggl(f'^2+{2s_f^2\over r^2}\biggr)d^3r,
\eqno(16) $$
kaon and pion masses $m_K,\;m\pi$ are taken from experiment.
The quantity $SC=<s_\nu^2>/2=<1-D_{88}^{(8)}>/3$ averaged over the baryon
$SU(3)$ wave function defines its strangeness content. Without configuration 
mixing, i.e. when flavor symmetry breaking terms in the lagrangian are 
considered as small perturbation, $<s_\nu^2>_0$ can be expressed simply in 
terms of the $SU(3)$ Clebsh-Gordan coefficients. The values of $<s_\nu^2>_0$
for the octet, decuplet, antidecuplet and some components of higher multiplets
are presented in {\bf Table 2}. In this approximation the components of $\{10\}$ 
and $\{\overline{10}\}$ are placed equidistantly, and splittings of decuplet
and antidecuplet are equal.

The spectrum of states with configuration mixing and diagonalization of the
hamiltonian in the next order of perturbation theory in $H_{SB}$ is given in 
{\bf Table 2} (the code for calculation was presented by H.Walliser).
The calculation results in the Skyrme model with only one adjustable parameter -
Skyrme constant $e$ ($F_\pi=186\,Mev$ - experimentally measured value) are shown
as variants A and B. The values of $<s_\nu^2>$ become lower when configuration mixing
takes place, and equidistant spacing of components inside of decuplet and especially 
antidecuplet is violated, see also {\bf Fig.2}.

It should be stressed here that the chiral soliton approach in its present state
can describe the differences of baryon or multibaryon masses \cite{7,8,10,19}.
The absolute values of mass are controlled by loop corrections of the order of
$N_C^0\sim 1$ which are estimated now for the case of $B=1$ only \cite{21}. 
Therefore, the value of nucleon mass in {\bf Table 2.} and {\bf Fig.2} is taken to be equal to 
the observed value.

\begin{center}
\begin{tabular}{|l|l|l|l|l|l|}
\hline
   &      & A & B& C &      \\
\hline
$\Theta_\pi \,(Gev^{-1})$ & --- &$6.175 $& $5.556$ & $5.61$ & -    \\
$\Theta_K  \,(Gev^{-1})$   & --- &$2.924 $  & $2.641$ & $2.84$ & -    \\
$\Gamma_{SB} \,\;(Gev) $       & --- & $1.369$  & $1.244 $ & $1.45$ & -    \\
\hline
\hline
$Baryon|N,Y,I,J>$ &$<s_\nu^2>_0$& A & B  & C   & $Data$\\
\hline
$\Lambda\,|8,0,0,1/2>$  &$0.60$ & 155&139& 164    &176 \\
$\Sigma\,|8,0,1,1/2> $  &$0.73$ & 263&243& 277    &254 \\
$\Xi  \,|8,-1,1/2,1/2>$ &$0.80$ & 371&335& 393 &379 \\
\hline
$\Delta \,|10,1,3/2,3/2>$ &$0.58$& 289&319& 314 &293 \\
$\Sigma^*|10,0,1,3/2>$  &$0.67$  & 418&433&  452 &446 \\
$\Xi^* |10,-1,1/2,3/2>$  &$0.75$ & 544&545&  586 &591 \\
$\Omega \,|10,-2,0,3/2> $ &$0.83$& 665&648& 715&733 \\
\hline
$\Theta^+\,|\overline{10},2,0,1/2>$&$0.50$&580&625 & 600&601 \\
$ N^*\,|\overline{10},1,1/2,1/2>  $&$0.58$&694&725 &722&771? \\
$\Sigma^*\,|\overline{10},0,1,1/2>$&$0.67$&792&810 & 825 &830? \\
$\Xi^{**}|\overline{10},-1,3/2,1/2>$&$0.75$&814&842& 847 &? \\
\hline
$\Theta^*\,|27,2,1,3/2>$ &$0.57$& 707  &758  & 750 &-  \\ 
$\Omega^*\,|27,-2,1,3/2>$ &$0.82$&989 & 1011  & 1048 &-  \\
\hline  
$ \;X\,|35,1,5/2,5/2>$ &$0.44$& 784  & 878     &  853 &-  \\

$ \;\;\;\,|35,-3,1/2,5/2>$ &$0.85$& 1269  & 1312 & 1367 &-  \\
\hline  
$ \;\,|28,2,3, 5/2>$       &$0.61$&  1938 & 2136 & 2043 &-  \\
$ \;\;\,|28,-4,0 ,5/2>$    &$0.78$& 2221  & 2379 & 2345 &-  \\
\hline
\end{tabular}
\end{center}

{\bf Table 2.} {\tenrm Values of masses of the octet, decuplet, antidecuplet and
some components of higher multiplets (with nucleon mass subtracted). 
A: $e=3.96$; B: $e=4.12$; C: fit with parameters $\Theta_K,\;\Theta_\pi$ and 
$\Gamma_{SB}$ \cite{10}, which are shown as well. }
\vglue 0.2cm
As it can be seen from {\bf Table 2}, the agreement with data for pure Skyrme model with 
one parameter is not so good, but the observed mass of $\Theta^+$ is reproduced
with some reservation. To get more reliable predictions for masses of other
exotic states the more phenomenological approach was used in \cite{10} where
the observed value $M_\Theta =1.54\,Gev$ was included into the fit, and 
$\Theta_K,\;\Gamma_{SB}$ were the variated parameters (variant C in {\bf Table 2} 
and {\bf Fig.2}). The position of some components of $\{27\},\;\{35\}$ and $\{28\}$
plets is shown as well.

\begin{figure}[h]
\label{spectrum}
\begin{center}
\epsfig{figure=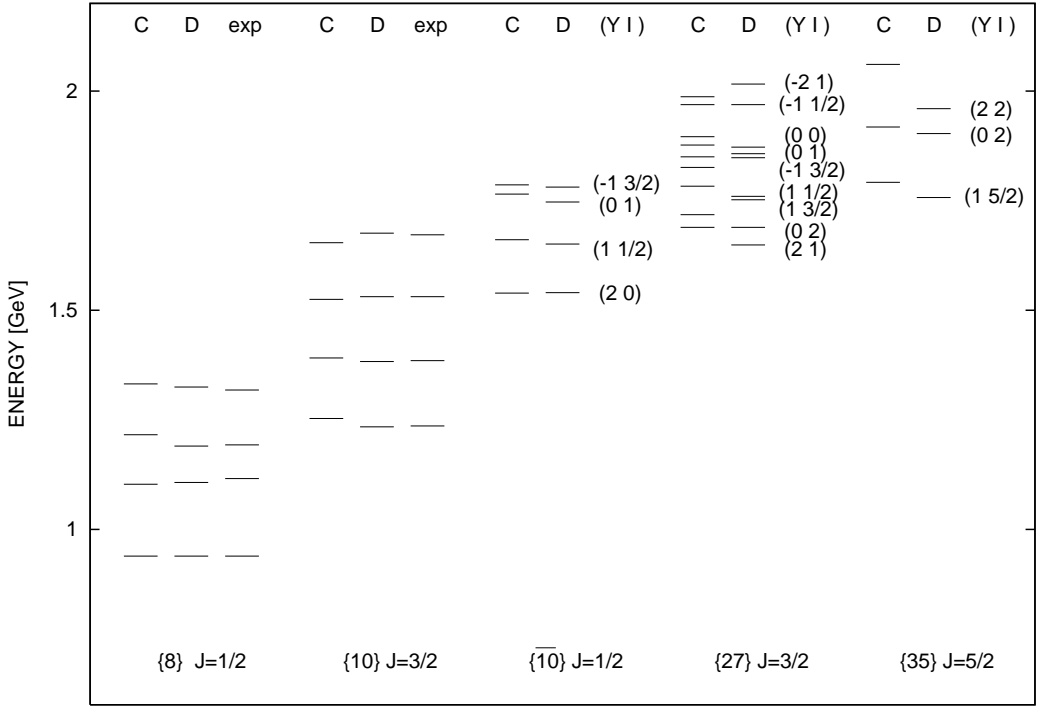,width=10cm,angle=270}
\protect\caption{Lowest rotational states in the $SU(3)$ 
soliton model for fits C and D. The experimental masses
of the $\{8\}$ and $\{10\}$ baryons are depicted for comparison.
Not all states of the $\{35\}$ are shown. This figure is taken from \cite{10}.}
\end{center}
\end{figure}
The variant D shown in {\bf Fig.2} takes into account the term in $H_{SB}$ which 
appears from the $\rho-\omega$ mixing in effective lagrangian \cite{7,10}:
$$ H_{SB}^{(2)} =\,-{\Delta\over \Theta_\pi}\sum_{a=1}^3 D_{8a}R_a \eqno (17)$$
The best description of the octet and decuplet masses was obtained at
$\Delta=0.4$. Such contribution was included also in \cite{8} where the linear
in hypercharge term $H_{SB}^Y=\beta Y$ with $\beta\simeq -156\,Mev$ plays an
important role. Such term is absent in approach \cite{7,10}.

It looks astonishing at first sight that the state $\Theta^+$ containing 
strange
antiquark is lighter than nonstrange component of antidecuplet, $N^*(I=1/2)$.
But it is easy to understand if we recall that all antidecuplet components contain
$q\bar{q}$ pair: $\Theta^+$ contains 4 light quarks and $\bar{s}$, $N^*$ 
contains 3 light quarks and $s\bar{s}$ pair with some weight, $\Sigma^* \in 
\{\bar{10}\}$ contains $u,d,s$ quarks and $s\bar{s}$, etc.

The mass splitting inside of decuplet is influenced essentially by its mixing
with $\{27\}$-plet components \cite{10}, see Fig.1, which increases this 
splitting considerably - the effect ignored in \cite{8}. The mixing of 
antidecuplet with the octet of baryons has some effect on the position of
$N^*$ and $\Sigma^*$, the position of $\Theta^* $ and $\Xi^*_{3/2}$ is influenced
by mixing with higher multiplets \cite{10}.

The component of $\{35\}$-plet with zero strangeness and $I=J=5/2$ is of special 
interest because it has the smallest strangeness content (or $s_\nu^2$) -
smaller than nucleon and $\Delta$. As a consequence of isospin conservation by
strong interactions it can decay into $\Delta\pi$, but not to $N\pi$ or $N\rho$.
According to the results presented in {\bf Table 2}, the components of $\{28\}$
plet containing 2 $q\bar{q}$ pairs, have the mass considerably greater than
that of other multiplets on Fig.1.

All baryonic states considered here are obtained by means of quantization of
soliton rotations in $SU(3)$ configuration space, and have therefore positive
parity. A qualitative discussion of the influence of other (nonzero) modes
- vibration, breathing - as well as references to corresponding papers can be 
found in \cite{10,16}. The realistic situation can be more complicated than 
somewhat simplified picture presented here, since each rotation state can have 
vibrational excitations with characteristic energy of hundreds of $Mev$.

If the matrix element of the decay $\Theta^+\to KN$ is written in a form
$$ M_{\Theta\to KN} = g_{\Theta KN}\bar{u}_N \gamma_5 u_\Theta \phi_K^\dagger
\eqno (18) $$
with $u_N$ and $u_\Theta$ - bispinors of final and initial baryons, then the
decay width equals to
$$\Gamma_{\Theta\to KN}=\frac{g^2_{\Theta KN}}{8\pi}
\frac{\Delta_M^2-m_K^2}{M^2} p_K^{cm} \eqno (19) $$
where $\Delta_M=M-m_N, \; M$ is the mass of decaying baryon, $p_K^{cm}$ - the
kaon momentum in the c.m. frame. For the decay constant we obtain then
$g_{\Theta KN} \simeq 4.4$ if we take the value $\Gamma_{\Theta\to KN}=10\,Mev$
as suggested by experimental data \cite{1}-\cite{5}. This should be compared 
with $g_{\pi NN}\simeq 13.5$. So, some suppression of the decay  
$\Theta\to KN$ takes place, but not 
large and understandable, according to \cite{8,22}.
\section{Exotic multibaryons}
There is no difference of principle, within the chiral soliton approach, 
between baryons and multibaryons, as it was demonstrated in previous sections. 
The latter are quantized configurations of
chiral fields which correspond to the minima of classical energy for arbitrary
baryon number. The equality between body-fixed isospin and spin of the 
quantized state, specific for hedgehog-type configuration, does not hold
anymore.

It is easily to understand that minimal (nonexotic) multiplets for $B=2$ coincide
with $m=1$ multiplets for $B=1$, i.e. they are antidecuplet,
including the deuteron - isosinglet state, $\{27\}$-plet, including the 
isotriplet $NN$-state (so called singlet deuteron), $\{35\}$ and $\{28\}$-
plets. Similarly, the minimal multiplets for $B=3$ are those for $B=1$ and
$m=2$, see {\bf Table 1}.

Here we show several examples of lowest exotic multiplets with $m=1$: the 
$\{\overline{35}\}$-plet 
for $B=2$, the $\{\overline{28}\}$-plet for $B=3$ and $\{\overline{80}\}$-plet
for $B=4$, {\bf Fig.3}.
There is isodoublet of positive strangeness dibaryons, $^2D_S^+,\;^2He_S^{++}$
\footnote{The chemical symbol is ascribed according to the total charge of
the baryonic state.}
with minimal quark contents $(\bar{s}\,3u\,4d),\;(\bar{s}\,4u\,3d)$,
which have the energy about $600\,Mev$ above $2N$-threshold, according to
calculation performed in \cite{20} in the slow rotator approximation.
The spectrum of all minimal dibaryons was calculated in \cite{20} as well.
 
\begin{figure}[h]
\label{multiplet}
\setlength{\unitlength}{1.0cm}
\begin{flushleft}
\begin{picture}(12,15)
\put(3,12){\vector(1,0){2.7}}
\put(3,12){\vector(0,1){4.5}}
\put(2.6,16.2){$Y$}
\put(5.4,11.6){$I_3$}
\put(2,9){$\{\overline {35}\}\, B=2$}
\put(1.9,15.1){$D_S^+ $}
\put(3.6,15.1){$He_S^{++}$}

\put(2.5,15){\circle*{0.2}}
\put(3.5,15){\circle*{0.2}}
\put(2,14){\circle*{0.12}}
\put(3,14){\circle*{0.1}}
\put(3,14){\circle {0.2}}
\put(4,14){\circle*{0.12}}

\put(1.5,13){\circle*{0.12}}
\put(2.5,13){\circle*{0.1}}
\put(2.5,13){\circle {0.2}}
\put(3.5,13){\circle*{0.1}}
\put(3.5,13){\circle {0.2}}
\put(4.5,13){\circle*{0.12}}

\put(1,12){\circle*{0.12}}
\put(2,12){\circle*{0.1}}
\put(2,12){\circle {0.2}}
\put(3,12){\circle*{0.1}}
\put(3,12){\circle {0.2}}
\put(4,12){\circle*{0.1}}
\put(4,12){\circle {0.2}}
\put(5,12){\circle*{0.12}}

\put(0.5,11){\circle*{0.2}}
\put(1.5,11){\circle*{0.1}}
\put(1.5,11){\circle {0.2}}
\put(2.5,11){\circle*{0.1}}
\put(2.5,11){\circle {0.2}}
\put(3.5,11){\circle*{0.1}}
\put(3.5,11){\circle {0.2}}
\put(4.5,11){\circle*{0.1}}
\put(4.5,11){\circle {0.2}}
\put(5.5,11){\circle*{0.2}}
\put(1,10){\circle*{0.2}}
\put(2,10){\circle*{0.2}}
\put(3,10){\circle*{0.2}}
\put(4,10){\circle*{0.2}}
\put(5,10){\circle*{0.2}}

\put(2.5,15){\line(1,0){1}}
\put(1,10){\line(1,0){4}}
\put(1,10){\line(-1,2){0.5}}
\put(0.5,11){\line(1,2){2}}
\put(5.5,11){\line(-1,2){2}}
\put(5.5,11){\line(-1,-2){0.5}}


\put(10,12){\vector(1,0){3}}
\put(10,12){\vector(0,1){4.5}}
\put(9.6,16.2){$Y$}
\put(12.7,11.6){$I_3$}
\put(9,9){$\{\overline{28}\}\,B=3 $}
\put(10.1,16.1){$He_S^{++}$}

\put(10,16){\circle*{0.2}}
\put(9.5,15){\circle*{0.12}}
\put(10.5,15){\circle*{0.12}}
\put(9,14){\circle*{0.12}}
\put(10,14){\circle*{0.12}}
\put(11,14){\circle*{0.12}}

\put(8.5,13){\circle*{0.12}}
\put(9.5,13){\circle*{0.12}}
\put(10.5,13){\circle*{0.12}}
\put(11.5,13){\circle*{0.12}}

\put(8,12){\circle*{0.12}}
\put(9,12){\circle*{0.12}}
\put(10,12){\circle*{0.12}}
\put(11,12){\circle*{0.12}}
\put(12,12){\circle*{0.12}}

\put(8.5,11){\circle*{0.12}}
\put(9.5,11){\circle*{0.12}}
\put(7.5,11){\circle* {0.12}}
\put(10.5,11){\circle*{0.12}}
\put(12.5,11){\circle*{0.12}}
\put(11.5,11){\circle*{0.12}}

\put(7,10){\circle*{0.2}}
\put(8,10){\circle*{0.2}}
\put(9,10){\circle*{0.2}}
\put(10,10){\circle*{0.2}}
\put(11,10){\circle*{0.2}}
\put(12,10){\circle*{0.2}}
\put(13,10){\circle*{0.2}}
\put(7,10){\line(1,2){3}}
\put(7,10){\line(1,0){6}}
\put(13,10){\line(-1,2){3}}



\put(6,3){\vector(1,0){4}}
\put(6,3){\vector(0,1){6}}
\put(6.1,8.7){$Y$}
\put(9.7,2.6){$I_3$}
\put(5,-1){$\{\overline{80}\}\,B=4 $}
\put(4.6,8.1){$He_S^{++}$}
\put(6.6,8.1){$Li_S^{+++}$}

\put(5.5,8){\circle*{0.2}}
\put(6.5,8){\circle*{0.2}}
\put(5,7){\circle*{0.12}}
\put(6,7){\circle*{0.12}}
\put(6,7){\circle {0.22}}
\put(7,7){\circle*{0.12}}

\put(4.5,6){\circle*{0.12}}
\put(5.5,6){\circle*{0.12}}
\put(5.5,6){\circle {0.22}}

\put(6.5,6){\circle*{0.12}}
\put(6.5,6){\circle {0.22}}
\put(7.5,6){\circle*{0.12}}

\put(4,5){\circle*{0.12}}
\put(5,5){\circle*{0.12}}
\put(5,5){\circle {0.22}}

\put(6,5){\circle*{0.12}}
\put(6,5){\circle {0.22}}

\put(7,5){\circle*{0.12}}
\put(7,5){\circle {0.22}}
\put(8,5){\circle*{0.12}}

\put(3.5,4){\circle*{0.12}}
\put(4.5,4){\circle*{0.12}}
\put(4.5,4){\circle {0.22}}

\put(5.5,4){\circle*{0.12}}
\put(5.5,4){\circle {0.22}}

\put(6.5,4){\circle*{0.12}}
\put(6.5,4){\circle {0.22}}
\put(7.5,4){\circle*{0.12}}
\put(7.5,4){\circle {0.22}}
\put(8.5,4){\circle*{0.12}}

\put(3,3){\circle*{0.12}}
\put(4,3){\circle*{0.12}}
\put(4,3){\circle {0.22}}
\put(5,3){\circle*{0.12}}
\put(5,3){\circle {0.22}}

\put(6,3){\circle*{0.12}}
\put(6,3){\circle {0.22}}

\put(7,3){\circle*{0.12}}
\put(7,3){\circle {0.22}}

\put(8,3){\circle*{0.12}}
\put(8,3){\circle {0.22}}
\put(9,3){\circle*{0.12}}
\put(2.5,2){\circle*{0.12}}
\put(3.5,2){\circle*{0.12}}
\put(3.5,2){\circle {0.22}}

\put(4.5,2){\circle*{0.12}}
\put(4.5,2){\circle {0.22}}

\put(5.5,2){\circle*{0.12}}
\put(5.5,2){\circle {0.22}}

\put(6.5,2){\circle*{0.12}}
\put(6.5,2){\circle {0.22}}

\put(8.5,2){\circle*{0.12}}
\put(8.5,2){\circle {0.22}}

\put(7.5,2){\circle*{0.12}}
\put(7.5,2){\circle {0.22}}

\put(9.5,2){\circle*{0.12}}

\put(2,1){\circle*{0.2}}
\put(3,1){\circle*{0.12}}
\put(3,1){\circle {0.22}}

\put(4,1){\circle*{0.12}}
\put(4,1){\circle {0.22}}

\put(5,1){\circle*{0.12}}
\put(5,1){\circle {0.22}}

\put(6,1){\circle*{0.12}}
\put(6,1){\circle {0.22}}

\put(7,1){\circle*{0.12}}
\put(7,1){\circle {0.22}}

\put(8,1){\circle*{0.12}}
\put(8,1){\circle {0.22}}

\put(9,1){\circle*{0.12}}

\put(9,1){\circle {0.22}}
\put(10,1){\circle*{0.2}}

\put(2.5,0){\circle*{0.2}}
\put(3.5,0){\circle*{0.2}}
\put(4.5,0){\circle*{0.2}}
\put(5.5,0){\circle*{0.2}}
\put(6.5,0){\circle*{0.2}}
\put(7.5,0){\circle*{0.2}}
\put(8.5,0){\circle*{0.2}}
\put(9.5,0){\circle*{0.2}}

\put(2,1){\line(1,2){3.5}}
\put(2,1){\line(1,-2){0.5}}
\put(2.5,0){\line(1,0){7}}
\put(5.5,8){\line(1,0){1}}
\put(10,1){\line(-1,2){3.5}}
\put(10,1){\line(-1,-2){0.5}}

\end{picture}
\vglue 1.0cm
\caption{\tenrm The $I_3-Y$ diagrams for the lowest exotic multibaryons: 
$\{\overline{35}\}$-plet of dibaryons, $\{\overline{28}\}$-plet of tribaryons 
and $\{\overline{80}\}$-plet of tetrabaryons, with $\;m=1$.
Large full circles show the exotic states, smaller - the cryptoexotic states.}

\end{flushleft}
\end{figure}
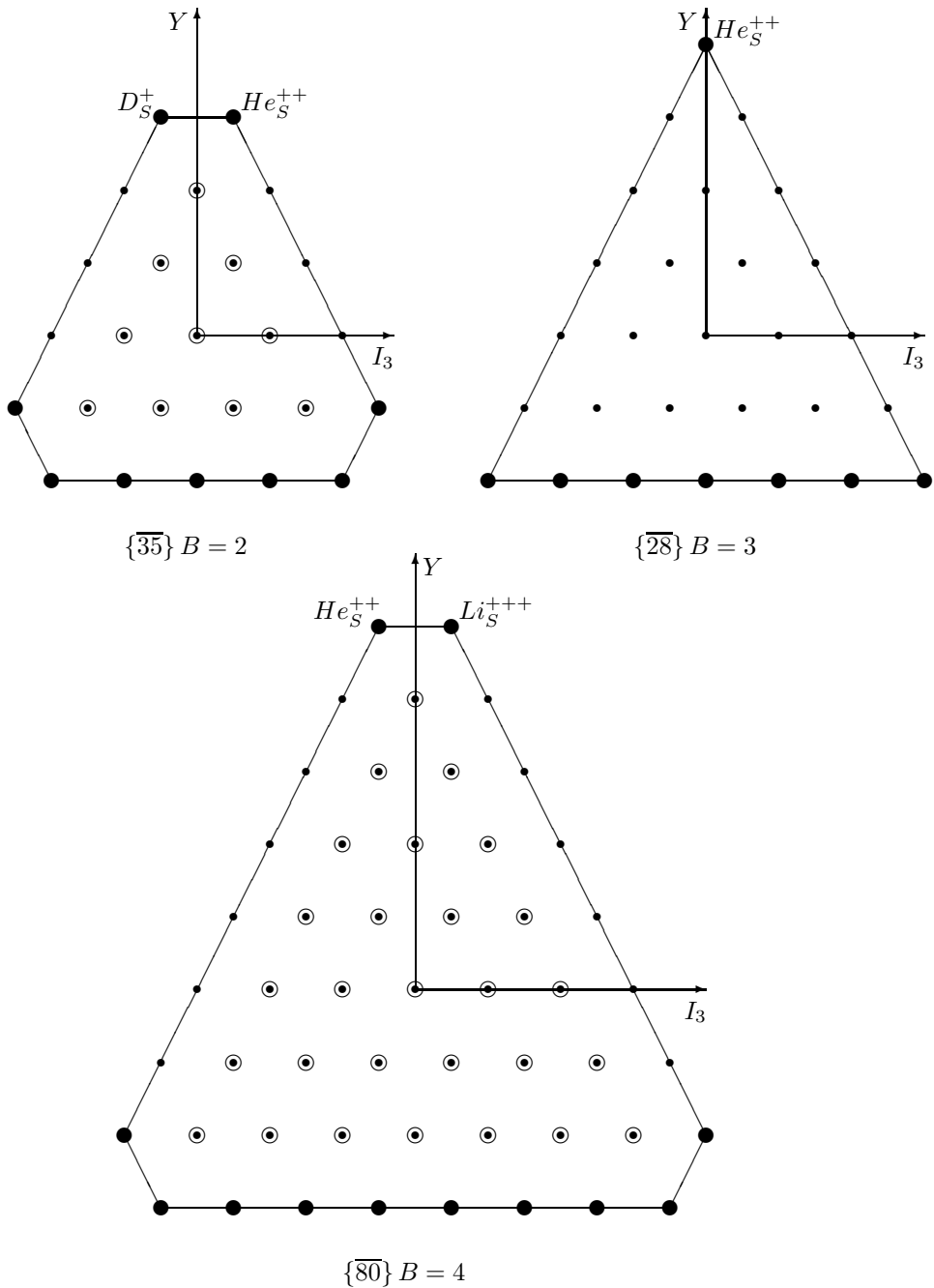

For $B=3$ there is positive strangeness tribaryon (isosinglet) $^3He_S^{++}$,
its quark content is $(\bar{s}\,5u\,5d)$. The position of the components of 
this multiplet is not calculated yet. One can state, however, in the spirit of
the version of the bound state model developed in \cite{23,24}, that the 
difference
of the masses of positive strangeness isosinglet and ground state of $^3He$
$$ M_{^3He_S}\, -\, M_{^3He} = \bar{\omega}_{S,B=3} + O(1/N_c) \eqno (20) $$
with $\bar{\omega_S}$ - the energy of antistrangeness excitation.
For $B=4$ there is positive strangeness isodublet $^4He_S^{++}$ - 
$^4Li_S^{+++}$ with minimal content $(\bar{s}\,6u\,7d)$ and $(\bar{s}\,7u\,6d)$.
Similarly, we have
$$ M_{^4He_S}\, -\, M_{^4He} = \bar{\omega}_{S,B=4} + O(1/N_c) \eqno (21) $$
The nonstrange components of such exotic multiplets (i.e. those with $Y=B$)
have the difference of masses
$$ M_{Y=B}\, -\, M_{B,ground\, st.} =\bar{\omega}_{S,B} + \omega_{S,B} +O(1/N_c),
\eqno (22) $$
and further $\omega_{S,B}$ should be added for each unit of strangeness, but the 
whole method \cite{23} works when strangeness is not large ($1-2$ units, not 
more).
The energies of flavor and antiflavor excitation for multiskyrmion were
calculated in \cite{24} for baryon numbers up to $22$, for $B\,>8$
within rational map approximation \cite{25}, using the results obtained in
comprehensive paper \cite{26}. 

Their characteristic feature is that they depend slightly on $B$-number.
It is known that the difference between antiflavor and flavor excitation 
energies \cite{23}
$$ \bar{\omega}_{F,B} - \omega_{F,B} = N_CB/(4\Theta_{F,B}), \eqno(23)$$
for any flavor (strangeness, charm or beauty) and baryon number.
Since $\Theta_{F,B} \sim B$ roughly \cite{24}, this difference depends weakly 
on B-number and scales like $N_C^0\sim 1$ \cite{23}.
Numerically $\bar{\omega}_S$ is close to $600\,Mev$ with small variations
\cite{24}. However, the $1/N_C$ corrections are not negligible, and this 
question deserves further study.

The qualitative treatment becomes very easy when the kaon mass is large enough.
In this case one obtains \cite{24}
$$\omega_{S,B} \simeq {\tilde{m}_K\over 2} \sqrt{\frac{\Sigma_B}{\Theta_{K,B}}}
 -  {3B\over 8\Theta_{K,B}} \eqno (24) $$

$$ \tilde{\omega}_{S,B} \simeq {\tilde{m}_K\over 2} 
\sqrt{\frac{\Sigma_B}{\Theta_{K,B}}} +
 {3B\over 8\Theta_{K,B}}  \eqno (25) $$
with $\tilde{m}_K^2 = F_K^2 m_K^2/F_\pi^2 - m_\pi^2 $, $\Sigma_B$ and 
$\Theta_{K,B}$ are given by expressions similar to $(16)$ and $(12)$. 
The ratio $r_{K,B}=\Sigma_B /\Theta_{K,B}$ decreases slightly with 
increasing $B$,  it can be proved rigorously that $r_{K,B} < 4F_\pi^2/F_K^2$
\cite{24}, therefore we have always $\omega_K <\, m_K$, and strangeness is 
bound for any $B$-number, with slightly increasing binding.

For antistrangeness the treatment simplifies if $F_K=F_\pi$, and we take this 
equality for the moment. Then
$$ \tilde{\omega}_{K,B} \simeq {m_K\over 2} r_{K,B}^{1/2} + \frac{3B}
{8\Theta_{K,B}}. \eqno (26) $$
Numerically $r_{K,B}^{1/2}$ decreases from $1.53$ for $B=1$ to $1.48$ for $B=4$
and $3B/(8\Theta_{K,B})$ is about $180\,Mev$ \cite{24} for $e=4.12$, but really
the first ratio depends on $e$ very weakly. So, we have
$$ \tilde{\omega}_K \simeq 0.76 m_K + 180 \,Mev$$
for $B=1$ and very close relations for other $B \leq 4$.
Evidently, with increasing $m_K$ antistrangeness also becomes bound, similar
to strangeness (more precise, for $m_K > \sim 750\,Mev$).
Corrections $\sim 1/N_c$ and $F_K/F_\pi =1.22 $ increase the critical value 
of $m_K$.  These conclusions agree with those made recently in \cite{27}.

Anticharm and antibeauty have chances to be bound: we obtain the corresponding
excitation energies $\bar{\omega}_c \sim (1.75 -1.8)Gev$ for $B$ between $4$
and $1$, $F_D/F_\pi \simeq 1.5$, and for antibeauty $\bar{\omega}_b 
\sim (4.9 - 5.0)Gev $ for the ratio $F_b/F_\pi \sim 2$ \cite{24}. So, these 
energies are smaller than corresponding meson masses, but to make more 
definite conclusions the $\sim 1/N_c$ corrections should be treated carefully.

The positive strangeness dibaryons should decay into $KNN$, tribaryons - into
 $K3N$ final states, etc. with a width of same order of magnitude as 
$\Gamma_\Theta$.
There are also exotic states with negative strangeness: dibaryons with $S=-4$,
isospin $I=2$, with electric charge in the interval from $Q=-3$ to $Q=+1$,
and tribaryons with $S=-5$, $I=3$ and charge from $-4$ up to $+2$, see {\bf 
Fig.3}.
Tetrabaryons with $S=-7$ can have charge in the interval $-5$ to $+2$.
As usually, it would be difficult to produce such states (one of possibilities
are heavy ion collisions), but their detection
could be easier: they decay mainly into $\Xi$-hyperons and pions.
The large amount of exotic multibaryons looks embarrassing at first sight. One 
should keep in mind, however, that many of them are too broad (those which have
energy by some hundreds of $Mev$ above threshold) and can be hardly 
distinguishable from continuum.

To conclude this section, note that there are other predictions of states in
chiral soliton models which are exotic in the common meaning of this word:
for example, charmed or beautiful hypernuclei bound stronger than strange 
hypernuclei \cite{28}. The supernarrow electromagnetically decaying dibaryon
with width about $\sim 1 Kev$ below the $NN\pi$ threshold \cite{29} was 
observed in two experiments \cite{30,31}, but not confirmed in \cite{32} in
the mass interval below $1914\,Mev$. Its searches certainly deserve further
efforts.
\section{Conclusions and prospects} 
The mass and width of recently detected baryon with positive strangeness,
$\Theta^+$ are in agreement with predictions of the topological (chiral)
soliton model \cite{6,7,8} \footnote{As it was noted, from rigoristic point 
of view one could doubt in each of these predictions, therefore experimental
confirmation was necessary.}. Possibly, another exotic baryon with zero 
strangeness has been observed \cite{9}. To be sure that the observed
$\Theta^+$ belongs to antidecuplet, the measurement of its spin and parity
is necessary first of all, as well as establishing its partners in $SU(3)$ 
multiplet (antidecuplet).

The searches for the state $\Theta^* \in \{27\}$ with isospin $I=1$ are of
interest. The double charged state $\Theta^{*++}$ could appear as a resonance
in $K^+p$ system. Since this state is by $\sim (120-160)\,Mev$ heavier than
$\Theta^+$ \cite{10}, its width should be at least $3-4$ times greater than
that of $\Theta^+$. The absence of such resonance could be a serious problem 
for the whole chiral soliton approach.

Let us note also that the mass splitting inside of antidecuplet obtained in
\cite{7,10} is considerably smaller than in \cite{8} where it is about 
$540\,Mev$. In addition, the deviation from equidistant law is large in 
\cite{7,10,6} as a consequence of configuration mixing being taken into account.
As a result, the value of mass of the hyperon with isospin $I=3/2$, $\Xi^*_{3/2}$
obtained in \cite{10} is considerably smaller than in \cite{8}. It is worth
noting that its mass estimate made in \cite{22} within antiquark-diquark-diquark
model is close to our result \cite{10}. The value of the mass of $\Sigma^*\in
\{\overline{10}\}$ also is lower in \cite{10} and is more close to $\Sigma^*
(1770)$ than to $\Sigma^*(1880)$.

Many exotic resonances of interest have large values of isospin, therefore they
cannot be observed in reactions of pion or kaon scattering on nucleons, but
could be seen in reactions of two and more pions (kaons) production, similar to
reaction studied in \cite{9}. It could be attractive a possibility to identify
the state of mass $1.72\,Gev$ observed in \cite{9} with a component of 
$\{35\}$-plet with $S=0,\;I=5/2$. But the isospin selection rule for reaction of
electroproduction \cite{9} with one-photon exchange makes such identification 
difficult. Another possibility noted already in \cite{9}, is that it is 
cryptoexotic component of $\{27\}$-plet with isospin $3/2$ and mass about 
$1.76\,Gev$, according to \cite{10}.

Of course, there is no contradiction between chiral soliton approach and the
quark (or quark-diquark, etc.) picture of baryons and baryon resonances, as 
it is stated in some papers. Both approaches are dual, the first one
describes baryons or baryonic systems from large enough distances and allows to
calculate such characteristics where the details of internal structure of 
baryons are not essential, one of such characteristics is just the mass of 
baryons.

The consequences of discovery of new baryon resonance are considered in several 
recent papers \cite{22},\cite{23},\cite{33}-\cite{35} and others, many of them have been reviewed and analysed in \cite{35}.
Hopefully, the results obtained in \cite{1}-\cite{5} and \cite{9} open new
interesting chapter in physics of baryon resonances, and its new pages can
be devoted also to studies of baryonic systems with exotic properties, 
including (anti)charm and beauty quantum numbers.

I am thankful to H.Walliser for numerous conversations, the 
present talk is based to large extent on the paper \cite{10}. I'm
indebted also to B.O.Kerbikov, A.E.Kudryavtsev, L.B.Okun' for useful questions
and discussions, and to K.-H.Glander, O.Hashimoto, T.Nakano, R.Schumacher for 
discussions during the Symposium.

The work has been supported by the Russian Foundation for Basic Research,
grant 01-02-16615.\\
\vglue 0.2cm

\end{document}